%% file: 0_main.tex
\documentclass{article}
\usepackage{iclr2027_conference,times}
\usepackage{enumitem}
\usepackage[T1]{fontenc}
\usepackage[utf8]{inputenc}
\usepackage{amsmath,amssymb,booktabs,tabularx}
\usepackage{graphicx}
\usepackage[most]{tcolorbox}
\usepackage{listings}
\usepackage{placeins}
\usepackage{multirow}
\usepackage{float}
\definecolor{promptblue}{HTML}{286A9D}
\newcommand{\systemprompt}[2]{\tcbinputlisting{enhanced,breakable,listing only,colback=black!2,colframe=promptblue!65!black,colbacktitle=promptblue!85!black,coltitle=white,fonttitle=\sffamily\bfseries\small,title={#1},title after break={#1},boxrule=0.5pt,arc=1.5mm,left=2mm,right=2mm,top=1.5mm,bottom=1.5mm,listing options={basicstyle=\fontsize{8.5}{11}\selectfont\ttfamily,breaklines=true,columns=fullflexible,keepspaces=true,showstringspaces=false,breakatwhitespace=true,breakindent=0pt,breakautoindent=false},listing file={prompts/typeset/#2.txt}}}
\usepackage{hyperref}
\usepackage{url}
\graphicspath{{figures/}}
\newcommand{\papertablestyle}{\small\setlength{\tabcolsep}{5pt}\renewcommand{\arraystretch}{1.12}}

\title{%
\raggedright
SWE-Clar: Can Coding Agents Clarify\\
Incomplete Client Requirements and\\
Deliver Greenfield Software?
}

\iclrfinalcopy
\author{%
Jialei Li$^{1,2*\ddagger}$, Dingxian Wang$^{3*\dagger}$, Xiaoyan Zhao$^{2\dagger}$, Yimeng Bai$^{1,2\ddagger}$, Xuanyi Zhu$^{1,2\ddagger}$,\\
\bfseries Yang Zhang$^{2}$, Chao Yin$^{3}$, Wentao Shi$^{1,2\ddagger}$, Ganesh Prabhu$^{3}$,\\
\bfseries Karthik Gomadam Rajagopal$^{3}$, Brett Levert$^{3}$, Darvin Yi$^{3}$,\\
\bfseries Andrew Rabinovich$^{3}$, Zhao Chen$^{3}$\\[4pt]
\normalfont $^{1}$University of Science and Technology of China\quad $^{2}$National University of Singapore\\
\normalfont $^{3}$Upwork Inc.}
\hypersetup{hidelinks,pdftitle={SWE-Clar: Can Coding Agents Clarify Incomplete Client Requirements and Deliver Greenfield Software?},pdfauthor={Jialei Li, Dingxian Wang, Xiaoyan Zhao, Yimeng Bai, Xuanyi Zhu, Yang Zhang, Chao Yin, Wentao Shi, Ganesh Prabhu, Karthik Gomadam Rajagopal, Brett Levert, Darvin Yi, Andrew Rabinovich, Zhao Chen}}
\begin{document}
\maketitle
\lhead{Preprint. Under review.}
\begingroup
\renewcommand{\thefootnote}{}
\footnotetext{$^{*}$Equal contribution. Email: \texttt{jialeili@mail.ustc.edu.cn}}
\footnotetext{$^{\dagger}$Corresponding authors. Email: \texttt{dingxianwang@upwork.com}, \texttt{xzhao@nus.edu.sg}}
\footnotetext{$^{\ddagger}$Work done in collaboration with the National University of Singapore.}
\endgroup
\begin{abstract}
\input{1_abstract}
\end{abstract}
\input{2_introduction}
\input{3_benchmark}
\input{4_experiments}
\input{5_related_work}
\input{7_conclusion}
\input{8_ai_statement}
\bibliographystyle{iclr2027_conference}
\bibliography{9_reference}
\clearpage
\appendix
\input{10_appendix}
\end{document}

%% file: 1_abstract.tex
Greenfield software development in freelance settings requires coding agents to turn incomplete client requests into working software. This makes requirement clarification essential, yet clients may be unable to provide continuous feedback over long development cycles, and receiving their answers does not guarantee correct understanding of complex requirements. These challenges motivate two benchmark design choices: clarification precedes code implementation, and its effectiveness is assessed through explicit tests of requirement understanding. Guided by these choices, we introduce SWE-Clar, a benchmark grounded in real, high-value client commissions to evaluate requirement clarification and code implementation in this setting. It comprises 100 tasks based on real software development jobs on Upwork, each with at least US\$4,000 in total payments. Each task begins with multi-round interaction between a coding agent and a client agent, followed by separate runs for code implementation with acceptance testing and clarification assessment. For the latter, we construct 512 questions paired with reference answers from real client--freelancer interactions, covering both explicitly stated requirements and those inferred from the conversational context. Both runs begin from the same clarification context but remain isolated, preventing assessment questions from guiding code implementation. Experiments with nine frontier models show that even GPT-6 Astra, which has the highest observed implementation score, achieves a task pass rate of only 67\%, underscoring the challenge of independent software delivery in this setting. The dataset and evaluation code are available at \url{https://github.com/upwork/SWE-Clar}.

%% file: 2_introduction.tex
\section{Introduction}
\label{sec:introduction}

\begin{figure}[!t]
\centering
\includegraphics[width=0.98\linewidth]{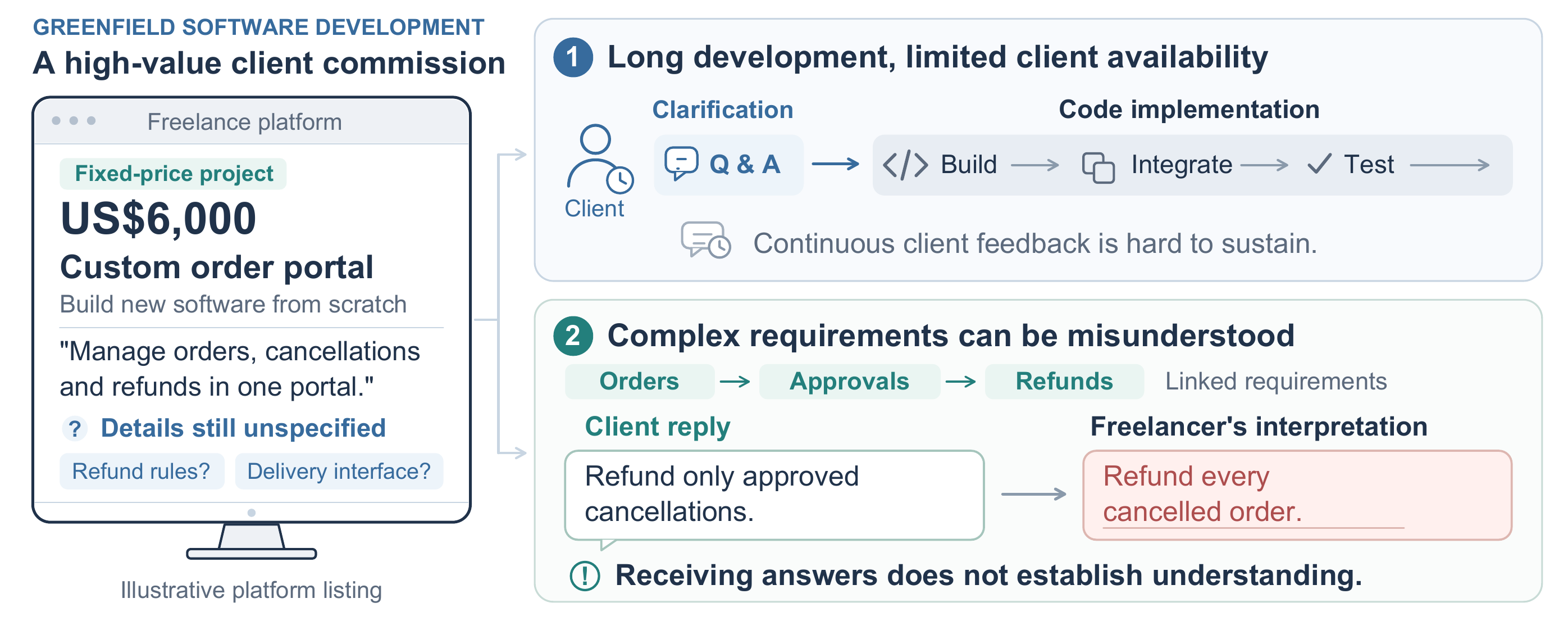}
\caption{A fictional commission illustrates the motivation for SWE-Clar in greenfield software development in freelance settings: (1) long development makes continuous client feedback difficult to sustain; (2) complex requirements can be misunderstood even after receiving replies.}
\label{fig:freelance_background}
\end{figure}

As large language models (LLMs) improve at coding and reasoning, coding agents are expanding from generating simple code snippets~\citep{chen2021codex} to tackling complex software engineering tasks~\citep{yang2024sweagent}. These advances create opportunities to automate software development services and deliver business value~\citep{peng2023impact,miserendino2025swelancer,yi2025upbench,mazeika2025rli}, particularly in greenfield software development, which involves building new software from scratch rather than modifying existing codebases~\citep{ding2025nl2repo,shi2026hypertau}. On freelance platforms such as Upwork\footnote{\url{https://www.upwork.com/}}, agents could act as freelancers to undertake high-value client commissions (Figure~\ref{fig:freelance_background}). However, these commissions typically begin with requests that leave critical functional and interface requirements unspecified~\citep{nuseibeh2000requirements}, so coding agents must proactively seek clarification through client interaction~\citep{mu2024clarifygpt}. Evaluating coding agents for greenfield software development in freelance settings therefore requires assessing both code implementation and requirement clarification.

Recent work has extended coding-agent evaluation beyond code implementation to include requirement clarification, enabling coding agents to interact with client agents to recover missing requirements~\citep{wu2025humanevalcomm,clarifycodebench2026,vijayvargiya2026ambig,icae2026,together2026,interact2026,king2026dialogue}. However, these benchmarks primarily focus on tasks derived from programming datasets and software repositories, leaving the demands of greenfield software development in freelance settings insufficiently represented. Our observations of real commissions for such greenfield software development projects highlight two characteristics, illustrated in Figure~\ref{fig:freelance_background}. First, building complete software projects from scratch often entails a prolonged development process, making continuous client feedback during code implementation difficult to sustain. Second, these projects involve multiple complex requirements that freelancers may not fully understand even after receiving answers to their questions. Consequently, benchmarks that assume continuous client feedback~\citep{together2026,interact2026} and assess clarification primarily through information acquisition~\citep{clarifycodebench2026,vijayvargiya2026ambig,icae2026,hil2026} may not adequately assess coding agents' ability to fulfill these commissions independently.

These characteristics motivate two design principles for conducting and evaluating requirement clarification. First, clarification should precede code implementation: the coding agent should complete its interaction with the client agent before implementing the software without further guidance. Second, clarification should be assessed using multiple requirement questions paired with reference answers. The coding agent's responses should be scored against these reference answers to assess requirement understanding rather than information acquisition alone. This assessment should remain separate from code implementation to avoid providing additional implementation guidance.

Guided by these design choices, we introduce \textbf{SWE-Clar}, a benchmark of 100 greenfield software development tasks based on real software development jobs on Upwork, each with at least US\$4,000 in total payments. For each task, a coding agent first engages in multi-round requirement clarification with a client agent. The resulting context then initializes two independent branches: one performs code implementation followed by acceptance testing, while the other evaluates clarification by scoring the coding agent's answers to requirement questions. To construct these questions, we use GPT-6 Astra to analyze recorded interactions between real clients and freelancers and identify requirements critical to successful software delivery, including those explicitly stated and those inferred from the broader conversational context. These requirements form the basis of 512 questions paired with reference answers, averaging approximately five per task. Together, these designs establish SWE-Clar as a benchmark grounded in real, high-value client commissions to evaluate requirement clarification and code implementation for greenfield software development in freelance settings. We evaluate nine frontier models on SWE-Clar to examine requirement clarification and independent code implementation in this setting. GPT-6 Astra has the highest observed implementation score, with a task pass rate of only 67\%, underscoring SWE-Clar's value as a challenging testbed for identifying limitations in coding agents' ability to fulfill real client commissions.

\begin{itemize}[leftmargin=*]
    \item We distill two design principles for requirement clarification in greenfield software development in freelance settings: clarification before code implementation and independent assessment of requirement understanding through question answering.

    \item We introduce SWE-Clar, a benchmark grounded in real, high-value client commissions to evaluate requirement clarification and code implementation in this setting, and release its dataset and evaluation code.

    \item We evaluate nine frontier models, demonstrating that independently completing these greenfield projects remains challenging even for the strongest evaluated model.
\end{itemize}

%% file: 3_benchmark.tex
\section{SWE-Clar: Benchmark Design}
\label{sec:benchmark}
In this section, we introduce SWE-Clar, then describe requirement clarification and the two independent evaluation branches. We subsequently explain how the tasks and evaluation materials are constructed, and examine the dataset and its differences from representative benchmarks.

\subsection{Overview}
\label{sec:overview}
\label{sec:setting}
We aim to evaluate requirement clarification and code implementation for greenfield software development in freelance settings. A coding agent must both understand what a client wants and deliver software that meets those requirements. Following the design principles in Section~\ref{sec:introduction}, SWE-Clar therefore places client interaction before code implementation and assesses the agent's understanding separately from the software it produces.

\begin{figure}[t]
\centering
\includegraphics[width=\linewidth]{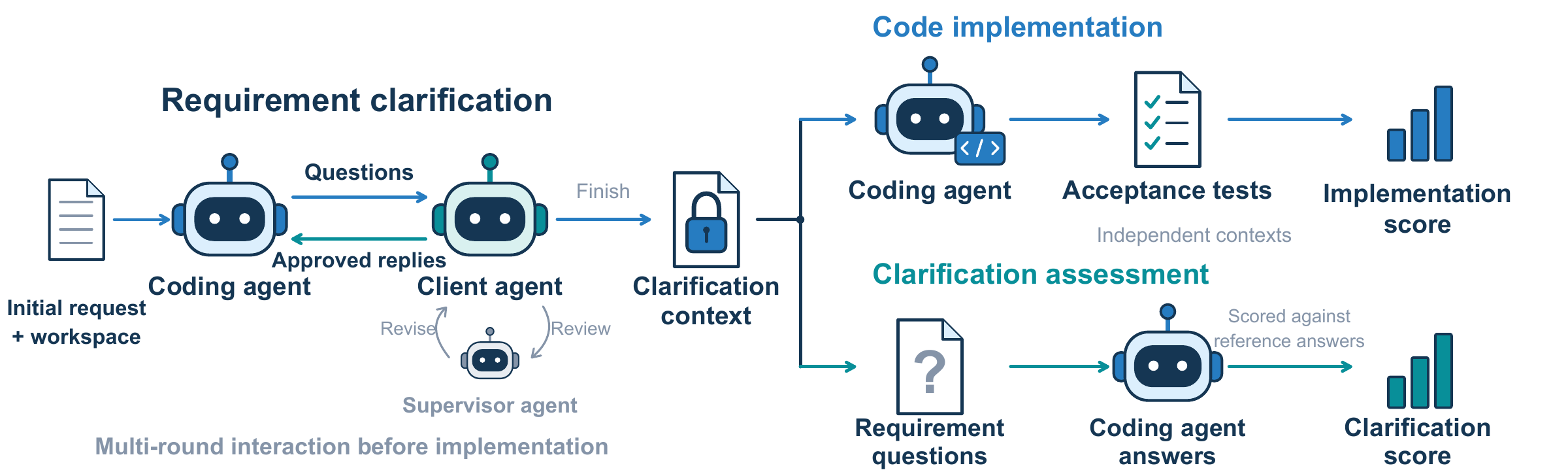}
\caption{SWE-Clar separates client interaction from two independent evaluations. The coding agent interacts with a client agent whose replies are reviewed by a supervisor agent. The resulting clarification context initializes both branches. Acceptance tests determine the implementation score; answers to requirement questions determine the clarification score. Assessment questions, answers, and scores do not enter the implementation branch.}
\label{fig:protocol}
\end{figure}

Figure~\ref{fig:protocol} shows the resulting workflow. Each task begins with an incomplete request for new software and a development workspace. A coding agent acts as the freelancer, while a separate client agent simulates the customer and answers from the complete requirements. The coding agent asks questions until it is ready to proceed. Its clarification context then initializes two independent runs: one implements the software and is scored through acceptance tests; the other answers requirement questions and is scored against reference answers. These runs yield the implementation score and clarification score, respectively, without further client guidance. The tasks are based on real software development jobs on Upwork, as detailed in Section~\ref{sec:construction}.

\suppressfloats[t]

\subsection{Requirement Clarification}
\label{sec:protocol}
The clarification phase determines what information the coding agent will have when it starts development. The agent inspects the supplied materials and asks the client agent about unclear or missing requirements. Its instructions explicitly require clarification of critical delivery details, such as how the software should be invoked and its expected inputs and outputs. The client agent responds using the complete requirements, confirming or correcting the agent's proposed interpretations and answering broader questions within their requested scope.

Because client replies guide both subsequent evaluations, we check them before they reach the coding agent. A supervisor agent reviews each draft for accuracy, completeness relative to the question, and whether it stays within the requested scope. When a draft needs revision, the supervisor explains the problem to the client agent, which revises the reply and submits it for review again. Only the approved reply reaches the coding agent; drafts and review feedback remain hidden.

The coding agent uses each approved reply to decide what to inspect or ask next, or whether to end clarification. One round consists of a question and its approved reply, regardless of the number of internal revisions. The agent may inspect the workspace during this phase but must wait until clarification ends to implement the software. This gives it multiple opportunities to resolve missing details early, while requiring subsequent development to proceed without further client guidance.

\subsection{Independent Evaluation}
\label{sec:metrics}
Both evaluations should start from the information available when clarification ends. However, an evaluation question may itself reveal a missing requirement; asking it in the same run before implementation could therefore change the software the agent produces. We avoid this by freezing the clarification context---the initial request, approved conversation, and information obtained from inspecting the workspace---and using it to start two independent runs. Their later messages and results remain separate, preventing assessment questions from guiding code implementation.

\paragraph{Implementation score.}
The first run tests whether the coding agent can turn the information it acquired into working software. The agent edits files, runs its own checks, and submits the implementation without access to further client replies, private acceptance tests, or reference implementations. Acceptance tests then run the submitted software and check its behavior. A task receives 1 if all tests pass and 0 otherwise. The implementation score is the percentage of tasks that pass, equivalent to task pass rate, and serves as our primary outcome.

\paragraph{Clarification score.}
The second run tests whether the coding agent can correctly explain the requested behavior. The agent answers requirement questions using the same clarification context and read-only task materials. It cannot execute commands, modify files, ask further client questions, or see reference answers. A judge agent compares its answers with the reference answers, without using the implementation's test results. Correct answers establish all material requested behavior without unresolved material contradictions and receive 1. Partial answers establish substantive correct behavior but leave a material omission or uncertainty, without an explicit material contradiction, and receive 0.5. Incorrect answers contain a material contradiction or provide no substantive correct answer and receive 0. Equivalent wording and correct inferences are accepted. The clarification score (\%) averages question scores within each task, then equally across tasks.

These scores separate the agent's ability to explain the requirements from its ability to implement them. For additional analysis, we also link each question to acceptance checks for the same requirement. Appendix~\ref{app:scoring} defines these requirement units (RUs), the auxiliary RU pass rate, and the answer-grading rubric. Table~\ref{tab:role_context} in Appendix~\ref{app:roles} summarizes information access and permitted actions; Appendix~\ref{app:prompts} provides the role prompts.

\subsection{Benchmark Construction}
\label{sec:construction}
The workflow requires tasks for which the initial request is incomplete but the intended software can still be specified and tested. We construct 100 such tasks based on real software development jobs on Upwork, each with at least US\$4,000 in total payments. For each commission, GPT-6 Astra analyzes the initial request and recorded client--freelancer conversations to identify requirements critical to delivery, including explicit statements and requirements inferred from context. We preserve the original business goal while replacing dependencies on client accounts and external deployment, allowing local development and evaluation. Appendix~\ref{app:source} details the task construction process.

The requirements identified in this process determine what the client agent knows. The coding agent receives an incomplete initial request and a workspace containing permitted files, data, and the runtime environment. These materials provide partial requirement information, not a complete checklist; the agent clarifies remaining details through inspection and client interaction. The complete requirements remain private and guide client replies, keeping the task fixed throughout the interaction.

The same requirements also provide the basis for both evaluations. We construct 512 questions paired with reference answers, covering the required behavior and its relevant conditions and exceptions. Separately, acceptance tests check whether the delivered software satisfies those requirements. For the illustrative commission in Figure~\ref{fig:freelance_background}, a question could ask which cancellations qualify for a refund, while tests would check the program's handling of approved and unapproved cancellations. The question checks the agent's understanding; the tests check the software's behavior.

Running the same tests on different implementations requires agreed delivery interfaces: how to invoke the software, what inputs it accepts, and what outputs it produces. We fix these interfaces together with the questions, reference answers, and tests before evaluation. The coding agent obtains the necessary interface details from the task materials and client interaction, while remaining free to choose the internal implementation.

Two reviewers with relevant industry backgrounds cross-checked the tasks for consistency among client requirements, requirement questions, and acceptance tests.

\subsection{Dataset Analysis and Discussion}
\label{sec:dataset_analysis}
\begin{figure}[htbp]
\centering
\begin{minipage}[t]{0.49\linewidth}
\centering
\vspace{0pt}
\includegraphics[width=\linewidth]{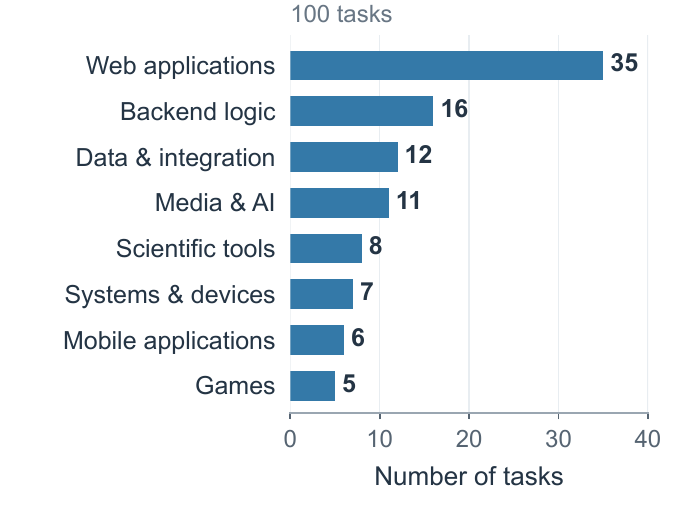}
\caption{The 100 tasks grouped by development type. Bars are ordered by task count, with counts shown at their ends.}
\label{fig:benchmark_task_distribution}
\end{minipage}\hfill
\begin{minipage}[t]{0.49\linewidth}
\centering
\vspace{0pt}
\includegraphics[width=\linewidth]{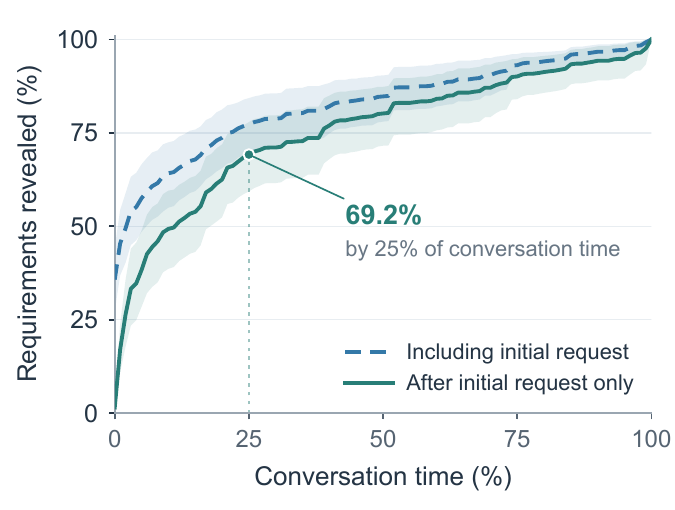}
\caption{Requirements revealed as client conversations progress. Time runs from the first to the last recorded message. Shading shows pointwise 95\% confidence intervals from conversation-level bootstrap resampling.}
\label{fig:source_disclosure}
\end{minipage}
\end{figure}

\paragraph{Task coverage.}
The 100 tasks span eight categories of development content and several runtime environments, with 512 requirement questions and 3--7 questions per task. Figure~\ref{fig:benchmark_task_distribution} shows the number of tasks in each category. This variety allows SWE-Clar to examine clarification and implementation across different types of greenfield projects.

\paragraph{Source conversations.}
The source conversations also show when requirements missing from the initial request become explicit. We identify their first disclosure and normalize time between each conversation's first and last visible messages. Among conversations with later-disclosed requirements, an average of 69.2\% appear within the first quarter of the observed period (Figure~\ref{fig:source_disclosure}). This concentration supports examining early clarification, although conversation timestamps alone do not establish when implementation began or whether clients remained available afterward.

\paragraph{Comparison with existing benchmarks.}
Table~\ref{tab:benchmark_comparison} distinguishes SWE-Clar's task setting and evaluation from representative benchmarks. SWE-Lancer already evaluates paid freelance work, but its coding tasks modify an existing codebase without client clarification~\citep{miserendino2025swelancer}. ClarifyCodeBench and Ambig-SWE study clarification for programming problems and repository issues~\citep{clarifycodebench2026,vijayvargiya2026ambig}, while ICAE-Bench extends interaction to from-scratch repository generation~\citep{icae2026}. SWE-Together and SWE-Interact assess development with continued user feedback and report interaction or goal-coverage diagnostics~\citep{together2026,interact2026}. SWE-Clar instead focuses on independently completing real, high-value greenfield commissions after early client interaction. Its separate question-answering run directly checks which requirements the agent understands, while acceptance tests check what it delivers. Evaluating both from the same clarification context makes these outcomes comparable without letting assessment questions influence the software.

\begin{table}[htbp]
\centering\papertablestyle

\caption{Representative coding benchmarks compared by task setting, client interaction, and evaluation. Independent requirement QA scores the coding agent's answers in a context isolated from code implementation. SWE-Lancer refers to its individual-contributor coding tasks.}
\label{tab:benchmark_comparison}
\begin{tabularx}{\linewidth}{@{}>{\raggedright\arraybackslash}p{0.21\linewidth}*{3}{>{\raggedright\arraybackslash}X}@{}}
\toprule
\textbf{Benchmark} & \textbf{Task setting} & \textbf{Client interaction} & \textbf{Evaluation} \\
\midrule
SWE-Lancer\newline\citeyearpar{miserendino2025swelancer} & Upwork commissions; existing-codebase changes & No client clarification & End-to-end code tests \\
\addlinespace[2pt]
ClarifyCodeBench\newline\citeyearpar{clarifycodebench2026} & LiveCodeBench; programming problems & Questions before code generation & Question quality and efficiency; code tests \\
\addlinespace[2pt]
Ambig-SWE\newline\citeyearpar{vijayvargiya2026ambig} & SWE-bench Verified; issue resolution & Queries to a simulated issue reporter & Information recovery; patch tests \\
\addlinespace[2pt]
SWE-Together\newline\citeyearpar{together2026} & Recorded coding sessions; repository tasks & Feedback throughout development & Intent coverage, user correction, and repository correctness \\
\addlinespace[2pt]
SWE-Interact\newline\citeyearpar{interact2026} & Existing SWE benchmarks; repository tasks & Progressive disclosure and implementation feedback & Plan/code goal coverage; task verifiers \\
\addlinespace[2pt]
ICAE-Bench\newline\citeyearpar{icae2026} & Open-source repositories; from-scratch generation & Requirement queries during project building & Constraint coverage; black-box code tests \\
\midrule
\textbf{SWE-Clar} & High-value Upwork commissions; greenfield projects & Reviewed clarification before implementation & Independent requirement QA; code acceptance \\
\bottomrule
\end{tabularx}
\end{table}

%% file: 4_experiments.tex
\section{Experiments}
\label{sec:experiments}
In this section, we evaluate coding agents on SWE-Clar to answer three research questions:
\begin{itemize}[leftmargin=*]
    \item \textbf{RQ1:} How well do coding agents clarify requirements and implement greenfield software?
    \item \textbf{RQ2:} How much does clarification help when the initial requirements are incomplete?
    \item \textbf{RQ3:} How does requirement-answer correctness relate to implementation success?
\end{itemize}

\subsection{Experimental Setup}
\label{sec:experimental_setup}
\paragraph{Tasks and metrics.}
We assign each model the same 100 SWE-Clar tasks and report the two scores defined in Section~\ref{sec:metrics}: \textbf{implementation score}, the percentage of tasks that pass all acceptance tests, and \textbf{clarification score}, the average requirement-question score. Both are reported as percentages. Table~\ref{tab:main_results} presents the main comparison.

\paragraph{Models and protocol.}
We evaluate GPT-6 Astra, GPT-5.6 Sol, GPT-5.6 Luna, Claude Fable 5.1, Claude Opus 5, GLM-5.3, GLM-5.3 Flash, Kimi K3, and Qwen-3.8 Max as coding agents. We use the highest supported reasoning effort for each coding model. The standard protocol uses GPT-5.6 Luna as the client agent, supervisor agent, and judge agent. Each coding agent completes clarification before the two evaluation branches begin. Appendix~\ref{app:roles} describes role configuration and information access.

\paragraph{Comparisons and uncertainty.}
We first compare models under the standard protocol, then examine the effects of three requirement conditions and compare answers with code-test outcomes for the same requirements. For implementation-score differences between requirement conditions, we report 95\% confidence intervals (Appendix~\ref{app:controls}). Additional analyses examine completion cost, requirement disclosure across interaction rounds, and detailed answer--implementation outcomes and cases (Appendix~\ref{app:additional}), followed by clarification timing (Appendix~\ref{app:timing}) and reply quality before and after Supervisor review (Appendix~\ref{sec:validity}).

\FloatBarrier
\subsection{Overall Performance (RQ1)}
\label{sec:rq1}
We begin by examining how well the models clarify client requirements and complete the resulting software tasks. Table~\ref{tab:main_results} reports both scores. Two observations follow:
\begin{itemize}[leftmargin=*]
    \item \textbf{Complete software delivery remains challenging.} Implementation scores range from 39.0\% to 67.0\%. GPT-6 Astra achieves the highest observed score but still fails 33 of the 100 tasks.
    \item \textbf{Higher clarification scores accompany higher implementation scores.} Across the nine models, the two scores show a strong positive association (Pearson $r=0.89$). RQ3 examines this relationship further at the level of individual requirements.
\end{itemize}

\begin{table}[htbp]
\centering\papertablestyle

\caption{SWE-Clar evaluation on 100 tasks per model. Bold and underlined values indicate the highest and second-highest scores; higher is better. Brackets show 95\% task-bootstrap confidence intervals (10,000 resamples).}
\label{tab:main_results}
\begin{tabular}{@{}lrr@{}}
\toprule
Coding model & \shortstack{Implementation\\score (\%)} & \shortstack{Clarification\\score (\%)} \\
\midrule
GPT-6 Astra & \textbf{67.0} [58, 76] & \textbf{87.5} [85.0, 90.1] \\
Claude Opus 5 & \underline{63.0} [53, 72] & 84.0 [80.9, 86.8] \\
Claude Fable 5.1 & 60.0 [50, 69] & \underline{84.3} [81.0, 87.3] \\
GPT-5.6 Sol & 57.0 [47, 67] & 78.1 [74.9, 81.3] \\
Qwen-3.8 Max & 55.0 [45, 64] & 81.3 [77.9, 84.7] \\
GLM-5.3 & 54.0 [44, 64] & 76.8 [73.3, 80.2] \\
Kimi K3 & 49.0 [39, 59] & 76.9 [73.4, 80.3] \\
GPT-5.6 Luna & 49.0 [39, 59] & 71.8 [68.6, 74.9] \\
GLM-5.3 Flash & 39.0 [30, 49] & 73.2 [68.9, 77.3] \\
\bottomrule
\end{tabular}
\end{table}

\subsection{Effect of Clarification (RQ2)}
\label{sec:rq2}
We next examine how access to client requirements changes both scores. For GPT-5.6 Luna and GLM-5.3 Flash, we compare \emph{no client interaction}, the \emph{standard clarification protocol}, and \emph{complete requirements provided upfront}. The task workspace is retained, and complete requirements do not reveal private tests or reference implementations. Both models are compared on the same 100 tasks under each condition. Requirement answers in all three conditions are evaluated with the same grading rubric used in the main comparison. Figure~\ref{fig:information} presents the results; paired differences and intervals are reported in Appendix~\ref{app:controls}.

\begin{figure}[htbp]
\centering
\includegraphics[width=0.90\linewidth]{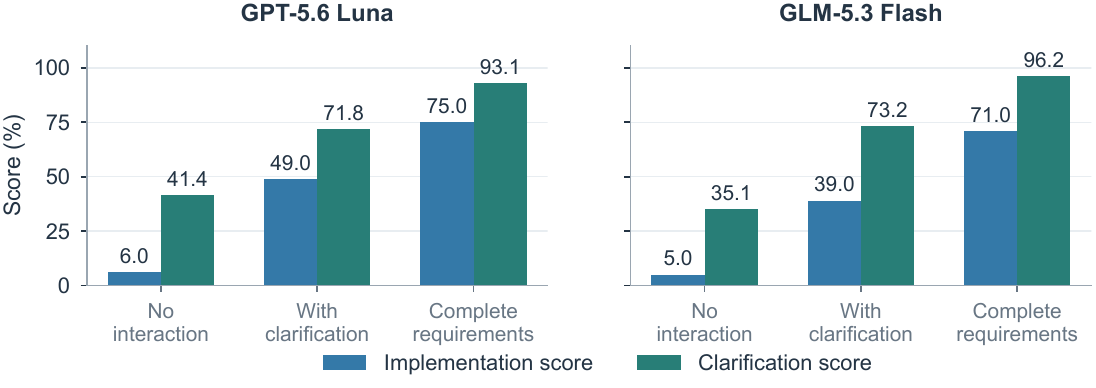}
\caption{Effect of requirement availability on the two scores. Each model is compared on 100 tasks under each condition. The same grading rubric is used across conditions and in the main comparison.}
\label{fig:information}
\end{figure}

\begin{itemize}[leftmargin=*]
    \item \textbf{Clarification improves both requirement understanding and software delivery.} With client interaction, GPT-5.6 Luna's implementation score rises from 6.0\% to 49.0\%, while GLM-5.3 Flash improves from 5.0\% to 39.0\%. Their clarification scores also increase by 30.4 and 38.1 percentage points, respectively. These results show improvements in both answering requirement questions and completing the software.
    \item \textbf{Clarification recovers only part of the gap to complete requirements.} Relative to no interaction, the standard protocol closes 62.3\% of the implementation-score gap for GPT-5.6 Luna and 51.5\% for GLM-5.3 Flash. Complete requirements further raise implementation scores to 75.0\% and 71.0\%, respectively. Missing information thus remains consequential, while the unsolved tasks under complete requirements show that implementation itself is also difficult.
\end{itemize}

\subsection{Requirement Answers and Implementation (RQ3)}
\label{sec:rq3}
\begin{figure}[htbp]
\centering
\includegraphics[width=\linewidth]{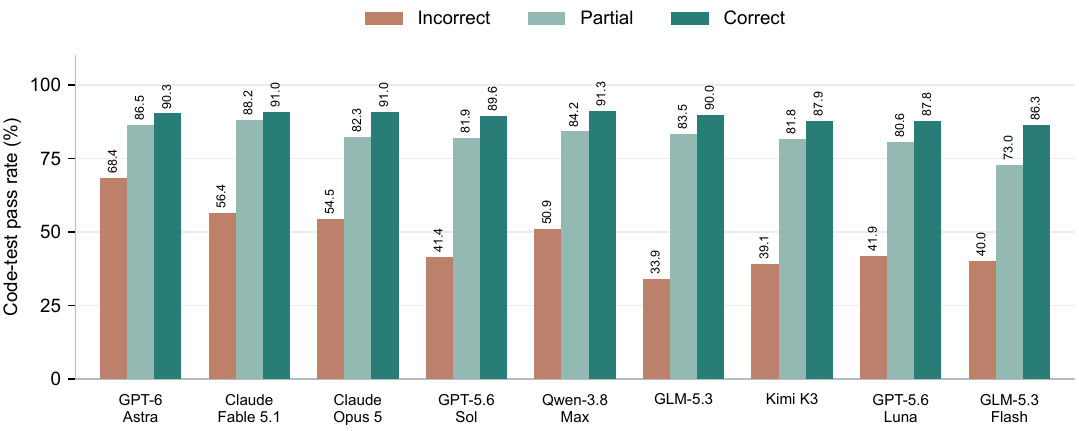}
\caption{Code-test pass rates within answer categories. Each bar uses the RUs in that model's answer category as its denominator. Incorrect includes contradictory or substantively unanswered requirements.}
\label{fig:agreement}
\end{figure}

The preceding results compare task-level performance across models. We next examine how requirement-answer correctness relates to implementation success. Each task contains multiple requirement units (RUs), each representing a specific client requirement with its own requirement question and acceptance tests (Appendix~\ref{app:scoring}). Each model is evaluated on all 512 RUs across 100 tasks. The joint analysis pairs the answer judgment for each of these 512 RUs with its corresponding code-test outcome. Figure~\ref{fig:agreement} shows results for correct, partial, and incorrect answers. Appendix~\ref{app:joint_results} provides the full joint distribution.

\paragraph{Incorrect answers identify requirements more likely to fail implementation.} Across models, code-test pass rates are 33.9--68.4\% for incorrect answers, compared with 86.3--91.3\% for correct answers and 73.0--88.2\% for partial answers. Correct answers have higher pass rates than partial answers in all nine models. Answer correctness is therefore associated with implementation success, while the two measurements capture different aspects of task performance.

\paragraph{Requirement answers and code tests provide complementary evidence.} The model-level correlation does not imply agreement on individual requirements: correct answers with failing implementations account for 6.1--7.8\% of all 512 RUs per model. Partial answers establish substantive correct behavior but leave a material omission or uncertainty without an explicit material contradiction. An unstated detail need not be absent from the code. For example, in the filtering case in Appendix~\ref{app:cases}, the answer omits the required empty-result behavior when no items match, but the implementation passes the corresponding test, which includes this case. Conversely, correct answers can accompany failing implementations, while incorrect answers can accompany passing implementations. Requirement questions thus measure what agents can explain, whereas acceptance tests check what they implement; the two outcomes are related but not interchangeable.

\paragraph{Disclosure does not guarantee correct answers.} After matching client-disclosure records to requirement answers from the same runs, 2.8--6.4\% of the initially hidden RUs fully disclosed by the end of clarification receive Incorrect labels across models (Appendix~\ref{app:disclosed_understanding}). This distinguishes acquiring requirement information from understanding it correctly.

\FloatBarrier

%% file: 5_related_work.tex
\section{Related Work}
\label{sec:related}

\paragraph{Software Development Benchmarks.}
Coding benchmarks have expanded from generating individual functions to completing software engineering tasks. HumanEval evaluates generated functions with executable tests~\citep{chen2021codex}, while SWE-bench asks agents to resolve GitHub issues in existing repositories~\citep{jimenez2024swebench}. SWE-Lancer further connects evaluation to paid freelance work through Upwork tasks from the Expensify codebase and their historical payouts~\citep{miserendino2025swelancer}. Another line of work evaluates complete repository generation: DevEval supplies product requirements and software designs, while NL2Repo-Bench provides natural-language specifications and evaluates generated repositories against upstream tests~\citep{li2025deveval,ding2025nl2repo}. These benchmarks establish the importance of evaluating working software in realistic settings. SWE-Clar focuses on greenfield software development in freelance settings, where agents must build new software from incomplete client requests. Its tasks preserve real business goals and require clarification before independent delivery.

\paragraph{Requirement Clarification.}
Incomplete requests require agents to determine what information is missing and obtain it before making consequential implementation choices. ClarifyGPT identifies ambiguous requirements and asks targeted questions to refine them before code generation~\citep{mu2024clarifygpt}. At the programming-problem level, HumanEvalComm and ClarifyCodeBench evaluate whether models ask useful clarification questions, with ClarifyCodeBench additionally measuring interaction efficiency~\citep{wu2025humanevalcomm,clarifycodebench2026}. At the repository level, Ambig-SWE examines how agents recover information omitted from issue descriptions~\citep{vijayvargiya2026ambig}. ICAE-Bench extends interactive evaluation to from-scratch repository generation, deriving incomplete requirements and grounded replies from open-source projects~\citep{icae2026}. SWE-Clar studies clarification in tasks based on real software development jobs on Upwork. Agents clarify requirements through multiple exchanges, then implement the software without further client guidance.

\paragraph{Interaction and Process Evaluation.}
Evaluating interactive agents involves both the user simulation and the evidence used to assess performance. $\tau$-bench combines simulated users with tool-using agents and checks task outcomes against the resulting environment state~\citep{yao2025taubench}. In coding, SWE-Together builds tasks from real user sessions and measures repository correctness and user correction, while its intent-coverage metric assesses the simulator's fidelity to the original user requests~\citep{together2026}. SWE-Interact provides feedback during development and examines goal coverage in plans and code revisions~\citep{interact2026}. SWE-RPG uses validated requirement and planning references to assess coding trajectories and attribute failures to requirement, planning, or coding stages~\citep{rpg2026}. These approaches show the value of examining more than the final code outcome. SWE-Clar obtains explicit evidence of requirement understanding through a separate question-answering run after clarification. Both runs start from the same context, allowing comparison without letting assessment questions guide implementation.

%% file: 7_conclusion.tex
\section{Conclusion}

In this work, we introduced \textbf{SWE-Clar}, a benchmark for requirement clarification and code implementation in greenfield software development in freelance settings. Based on real, high-value software development jobs on Upwork, its tasks require coding agents to clarify incomplete requests before independently developing software. We paired acceptance testing with a separate requirement-question assessment, yielding implementation and clarification scores without allowing assessment questions to guide coding. Evaluations of nine frontier models showed that complete delivery remains challenging. Comparisons across requirement conditions further showed that clarification improves performance but recovers only part of the gap to complete requirements, while model-level score correlations coexist with requirement-level answer--code disagreements, demonstrating the complementary information provided by the two evaluations.

Future work can extend SWE-Clar to changing requirements and different patterns of client availability, and investigate how coding agents can clarify client needs more effectively and carry them through to reliable software delivery.

%% file: 8_ai_statement.tex
\subsection*{AI use statement}

Generative AI tools assisted with manuscript translation and revision, literature research, interpretation of the reported results, and figure preparation, including plotting and layout code. GPT-6 Astra assisted with source-material processing, task construction, and development of requirement questions, acceptance tests, and reference implementations, with human cross-validation as described in Section~\ref{sec:construction} and Appendix~\ref{app:source}. LLM use in coding, client simulation, reply review, and answer grading is documented in the method, experiments, and appendix. We take full responsibility for verifying AI-assisted text, references, benchmark materials, and figures, and for all final claims and artifacts.

\subsection*{Reproducibility statement}
Our public repository (\url{https://github.com/upwork/SWE-Clar}) provides the 100 tasks, 512 requirement units, acceptance tests, role prompts, evaluation harness, environment build configurations, and reference, behaviorally equivalent, and targeted incorrect implementations for every task, together with offline validation tools. The appendix describes evaluation and scoring.

%% file: 10_appendix.tex
\FloatBarrier
\section{Task Construction}
\label{app:source}
\label{app:inventory}
Section~\ref{sec:construction} describes task construction. This appendix adds details on material handling and the analysis of requirement disclosure in source conversations.

We first identify 122 candidate jobs with at least eight milestones, at least US\$4,000 in total payments, and at least 20 messages. We select 100 jobs that can be reproduced locally, can avoid dependencies on external client accounts, and have deliverables suitable for evaluation through acceptance tests. Design-oriented tasks that are difficult to evaluate objectively are excluded. The monetary threshold refers to total payments for the source job.

GPT-6 Astra first processes the source materials. Two reviewers with relevant industry backgrounds then cross-check the tasks for consistency among client requirements, requirement questions, and acceptance tests. GPT-6 Astra and human contributors jointly develop the acceptance tests and reference implementations. All reference implementations pass every acceptance test for their respective tasks.

The supplied starter projects provide structure and, in some tasks, runnable baseline functionality; agents must complete the required target behavior. Greenfield refers to the new software projects underlying the tasks, rather than an empty starting directory. None of the 512 RUs across the 100 tasks passes its corresponding acceptance tests in the initial state, giving a starter RU pass rate of 0\%. Local workspaces use synthetic records and controlled services in place of original client accounts and deployment environments, allowing the agent to develop the required software locally. The release contains the curated benchmark tasks and evaluation harness, not original client conversations or identifying information.

We further check the acceptance tests using behaviorally equivalent implementations and targeted incorrect implementations. Across the 100 tasks, both reference and equivalent implementations pass all acceptance tests, while an incorrect implementation targeting one requirement per task is rejected by the corresponding RU tests. These checks verify acceptance of correct alternative implementations and detection of the tested requirement violations. The released repository includes the three implementation variants for every task, the requirement targeted by each incorrect implementation, and file hashes for integrity checks.

\subsection{Privacy Protection and Release Scope}
\label{app:privacy}
Real client commissions inform the business goals and requirements of the benchmark. The dataset and evaluation code are publicly released under the MIT License at \url{https://github.com/upwork/SWE-Clar}; the release excludes original conversations, client identities, source-job links, and conversation excerpts. Tasks use benchmark-specific identifiers without a public mapping to the original clients or commissions.

Tasks use synthetic records and controlled local services in place of original business records, client accounts, and deployment environments. The dataset contains no real client identities or contact information; fields needed for testing use synthetic data or placeholders. Evaluation does not require access to real client accounts. Before release, we inspect code, task materials, file paths, and media metadata for identifying information and credentials, while preserving third-party licenses and copyright notices. The paper reports aggregate statistics from source-conversation analysis without releasing individual conversation contents.

\subsection{Requirement Disclosure in Source Conversations}
\label{app:source_curves}
We analyze requirements in the initial request and client interaction, merge equivalent repetitions, and use the client's final confirmed requirements when earlier requirements change. We identify when these requirements become explicit in the conversation to analyze the timing of requirement disclosure. Unconfirmed freelancer proposals, scheduling, payment updates, and access logistics are excluded. The public dataset does not include supporting excerpts or message positions from the original conversations.

For each conversation, elapsed time is normalized between its first and last visible messages. Initial-request requirements are placed at time zero. The primary curve divides the cumulative number of later-disclosed requirements by the total number of such requirements observed in that conversation; conversations without later-disclosed requirements do not enter this mean. The supplementary curve includes initial-request requirements in both numerator and denominator. Figure~\ref{fig:source_disclosure} averages these curves equally across conversations; its shading shows pointwise 95\% confidence intervals obtained from 2,000 conversation-level bootstrap resamples.

\FloatBarrier
\section{Evaluation Protocol and Scoring}
\label{app:protocol}
\label{app:provenance}
\subsection{Roles, Context, and Tools}
\label{app:roles}
Our coding-agent runtime is built on Pi.\footnote{\url{https://github.com/earendil-works/pi}} Table~\ref{tab:role_context} specifies information access and permitted actions for the procedure in Sections~\ref{sec:protocol} and~\ref{sec:metrics}. The standard protocol sets no fixed limit on clarification rounds or question count. The coding agent autonomously decides when to end each phase: it calls \texttt{finish\_clarification} to end clarification and \texttt{submit\_delivery} to submit its implementation and end coding; Appendix~\ref{app:prompts} provides the full prompts.

The saved model configurations set reasoning effort to \texttt{max}. Where \texttt{max\_tokens} is unspecified, response length follows the corresponding service and model defaults rather than a common total-token budget. Acceptance tests directly determine implementation scores; the judge agent is used only to grade requirement answers.

\begin{table}[htbp]
\centering\papertablestyle
\caption{Agent responsibilities and information access. Both evaluation branches start from the frozen clarification context. Private requirement materials are accessible only to client simulation and evaluation agents.}
\label{tab:role_context}
\begin{tabularx}{\linewidth}{@{}>{\raggedright\arraybackslash}p{0.17\linewidth}>{\raggedright\arraybackslash}p{0.13\linewidth}>{\raggedright\arraybackslash}X>{\raggedright\arraybackslash}X@{}}
\toprule
Phase or branch & Agent & Available information & Actions and output \\
\midrule
Requirement clarification & Coding & Initial request, permitted workspace, approved conversation & Inspect materials, ask questions, decide when to stop \\
\addlinespace[2pt]
& Client & Complete requirements, task materials, approved conversation, current question & Draft or revise a reply within the requested scope \\
\addlinespace[2pt]
& Supervisor & Same information as the Client, plus the draft reply & Approve the reply or request revision \\
\addlinespace
Code implementation & Coding & Frozen context and development workspace & Implement, run development checks, submit software for acceptance testing \\
\addlinespace
Clarification assessment & Coding & Frozen context, requirement questions, read-only development materials & Answer questions without further interaction or implementation \\
\addlinespace[2pt]
& Judge agent & Reference requirements and answers, candidate answers, relevant acceptance tests & Grade requirement answers independently of code outcomes \\
\bottomrule
\end{tabularx}
\end{table}

\FloatBarrier
\subsection{Timeouts and Failure Handling}
\label{app:failure_handling}
The released harness defaults to a 900-second timeout without model or tool progress. This limits inactivity rather than total task duration. Coding has no total-duration limit by default; runtime configuration can override these limits.

Retryable execution errors are handled through saved workflow checkpoints, with at most five automatic retries at the same checkpoint. Exhausting this limit stops the run and records a failure. A retryable evaluation error triggers reevaluation of the submitted implementation without recoding. A completed delivery is not automatically retried because of a low score. Execution errors and acceptance-test failures are recorded separately.

\subsection{Requirement-Level Scoring}
\label{app:scoring}
Requirement units (RUs) link each client requirement to its question and acceptance tests. The auxiliary RU pass rate averages the proportion of passing RUs within each task and then equally across tasks.

For each RU, the judge agent evaluates the answers against the required behavior and agreed delivery interface. Corresponding acceptance tests clarify how these requirements translate into software behavior without introducing additional requirements. The judge examines all answers for the same task: relevant evidence in an answer to another RU also counts and need not be repeated in the target answer. Equivalent wording is accepted, but generic commitments such as ``follow the requirements'' do not establish knowledge of specific requirements. The judge evaluates the answers independently of code-test outcomes.

Requirement questions use consistent templates to identify the topic being assessed; answers must still specify the concrete behavioral rules in the reference requirements. Questions are provided only to the independent answer branch after the clarification context is frozen and do not enter the coding branch. This prevents the coding agent from inferring additional requirements from the assessment questions, preserving the evaluation of independent implementation based on the clarification context.

Requirement-answer grading uses Correct (1), Partial (0.5), and Incorrect (0). The judge evaluates answers against the requirements, delivery interface, and corresponding test source, without access to code-test outcomes; tests cannot introduce undeclared requirements. Verbatim evidence quotations are not required. Reference conflicts are resolved through repeated review or requirement-based manual adjudication, with the original records retained. Manual checks confirmed agreement between the judge's labels and human judgments on the reviewed samples.

\begin{table}[htbp]
\centering\papertablestyle
\caption{Requirement-answer grading rubric.}
\label{tab:rubric}
\begin{tabularx}{\linewidth}{@{}lXr@{}}
\toprule
Label & Criterion & Score \\
\midrule
Correct & States all material requirements correctly. & 1 \\
Partial & Establishes substantive correct behavior, with a material omission or uncertainty and no explicit material contradiction. & 0.5 \\
Incorrect & States incompatible behavior or provides no substantive correct answer. & 0 \\
\bottomrule
\end{tabularx}
\end{table}

\FloatBarrier
\section{Additional Experimental Results}
\label{app:additional}
This appendix supplements the main results with cost and RU pass rates, detailed comparisons of information conditions, disclosure during clarification, and full answer--implementation outcomes.

\subsection{Overall Performance and Cost}
\label{app:cost}
Cost covers the coding agent's calls during clarification and code implementation. Client, Supervisor, requirement-answering, and judge-agent calls are excluded. We compute cost from the saved input, output, and cache usage, priced at the actual recorded rates; we use the five-minute cache-write rate when the cache lifetime is unspecified.

\begin{figure}[htbp]
\centering
\includegraphics[width=0.84\linewidth]{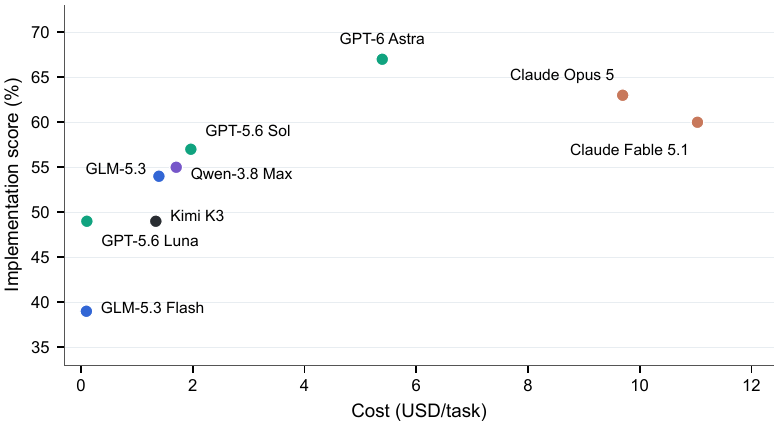}
\caption{Cost and implementation score. Cost includes coding-agent calls during clarification and implementation.}
\label{fig:cost_outcomes}
\end{figure}

Table~\ref{tab:auxiliary_results} summarizes mean cost and supplementary RU pass rate for 100 tasks per model. Figure~\ref{fig:cost_outcomes} compares estimated mean cost with the 100-task implementation scores.

\begin{table}[htbp]
\centering\papertablestyle
\caption{Supplementary RU pass rates and mean costs for 100 evaluation tasks per model.}
\label{tab:auxiliary_results}
\begin{tabular}{@{}lrr@{}}
\toprule
Model & RU pass (\%) & USD/task \\
\midrule
GPT-6 Astra & 88.5 & 5.391 \\
Claude Fable 5.1 & 86.7 & 11.033 \\
Claude Opus 5 & 87.2 & 9.694 \\
GPT-5.6 Sol & 84.4 & 1.965 \\
Qwen-3.8 Max & 85.9 & 1.703 \\
GLM-5.3 & 82.4 & 1.393 \\
Kimi K3 & 82.2 & 1.338 \\
GPT-5.6 Luna & 82.7 & 0.103 \\
GLM-5.3 Flash & 77.8 & 0.095 \\
\bottomrule
\end{tabular}
\end{table}

Higher call cost does not consistently correspond to higher task completion: Astra has a lower mean cost than either Claude model but a higher observed implementation score. Across models, RU pass rates exceed whole-task pass rates, distinguishing the ability to implement many individual requirements from the ability to satisfy all requirements of a task together.

\FloatBarrier
\subsection{Information Availability}
\label{app:controls}
For the information conditions in Section~\ref{sec:rq2}, Table~\ref{tab:information_controls} adds RU pass rates and exact scores, while Table~\ref{tab:information_differences} reports paired implementation-score differences. The 95\% confidence intervals use 10,000 task-level bootstrap resamples, selecting the same task IDs across conditions in each resample~\citep{efron1979bootstrap}.

\begin{table}[htbp]
\centering\papertablestyle
\caption{Information conditions on matched tasks within each candidate. Scores are percentages.}
\label{tab:information_controls}
\begin{tabular}{@{}llrrr@{}}
\toprule
Candidate & Information condition & \shortstack{Implementation\\score} & \shortstack{Clarification\\score} & \shortstack{RU pass\\rate} \\
\midrule
GPT-5.6 Luna & No client interaction & 6.0 & 41.4 & 37.5 \\
GPT-5.6 Luna & Standard protocol & 49.0 & 71.8 & 82.7 \\
GPT-5.6 Luna & Complete requirements & 75.0 & 93.1 & 93.2 \\
GLM-5.3 Flash & No client interaction & 5.0 & 35.1 & 36.0 \\
GLM-5.3 Flash & Standard protocol & 39.0 & 73.2 & 77.8 \\
GLM-5.3 Flash & Complete requirements & 71.0 & 96.2 & 91.1 \\
\bottomrule
\end{tabular}
\end{table}

\begin{table}[htbp]
\centering\papertablestyle
\caption{Paired task-pass-rate differences in percentage points, with task-bootstrap 95\% intervals.}
\label{tab:information_differences}
\begin{tabular}{@{}llrr@{}}
\toprule
Candidate & Comparison & Difference & 95\% interval \\
\midrule
GPT-5.6 Luna & Standard minus no interaction & 43.0 & [33.0, 53.0] \\
GPT-5.6 Luna & Complete minus standard & 26.0 & [14.0, 37.0] \\
GLM-5.3 Flash & Standard minus no interaction & 34.0 & [25.0, 43.0] \\
GLM-5.3 Flash & Complete minus standard & 32.0 & [20.0, 43.0] \\
\bottomrule
\end{tabular}
\end{table}

Both models improve from no interaction to the standard protocol and from the standard protocol to complete requirements; all four paired 95\% intervals exclude zero. Clarification therefore recovers part of the performance lost under incomplete initial information, while the further gains with complete requirements identify more thorough information acquisition as a promising direction.

\subsection{Requirement Disclosure during Clarification}
\label{app:disclosure}
Using the approved question--answer rounds defined in Section~\ref{sec:protocol}, we identify when each initially hidden RU is first fully disclosed through model-assisted annotations of client replies. Full disclosure requires all material clauses; partial and uncertain disclosures are not counted as complete, and repeated statements are counted once. The final value is carried forward after a conversation ends normally.

\begin{figure}[htbp]
\centering
\includegraphics[width=0.95\linewidth]{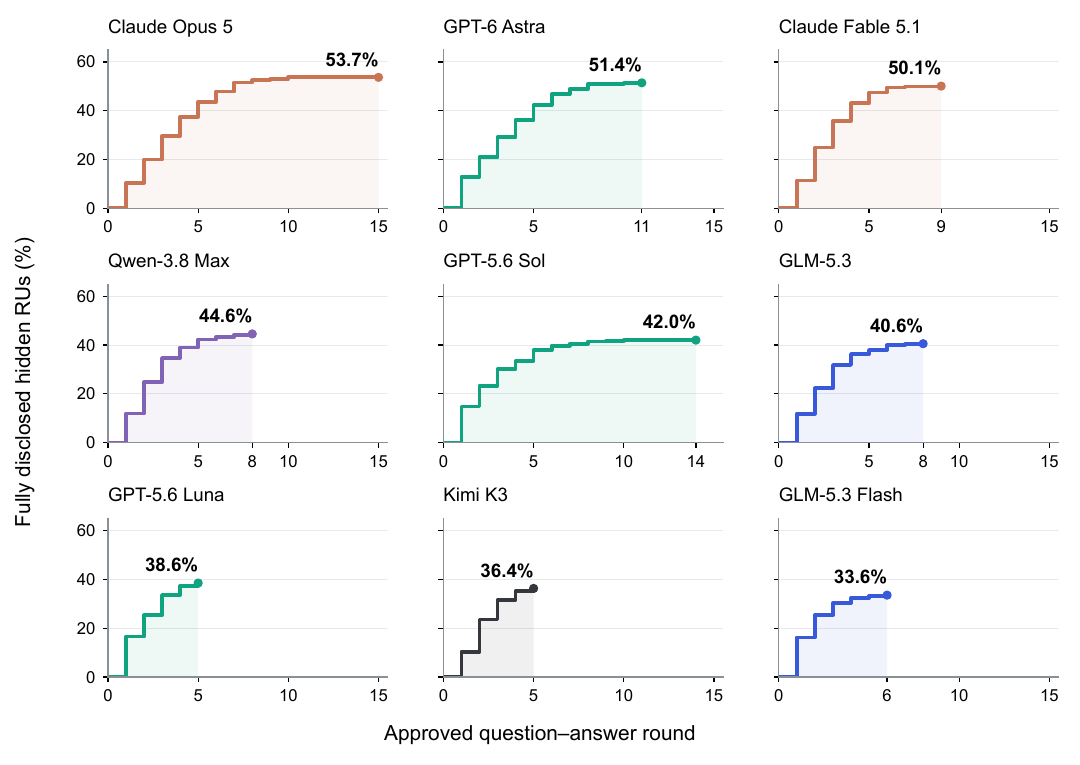}
\caption{Mean requirement disclosure after filtering conversations with outlying round counts. Models are ordered by their final proportion of fully disclosed hidden RUs. Additional horizontal-axis ticks mark each curve's final round.}
\label{fig:disclosure_appendix}
\end{figure}

To summarize disclosure in typical conversations, we exclude conversations whose question--answer counts fall below the first quartile minus 1.5 times the interquartile range or above the third quartile plus 1.5 times that range, separately for each model, and average disclosure fractions equally over the retained conversations. The curve records first complete disclosure, so later contradictions do not reduce it; each curve ends at the maximum retained round count for that model. Figure~\ref{fig:disclosure_appendix} shows the coding agent's progress in obtaining requirement information from the client, reflecting its ability to elicit requirements through clarification under the current client response policy. This ability to acquire requirement information through clarification represents a distinct capability from those measured by clarification and implementation scores.

After filtering, final mean disclosure ranges from 33.6\% to 53.7\% of initially hidden RUs across models. Under the current client response policy, agents thus end clarification with a substantial share of hidden requirements not yet fully disclosed, motivating further study of question coverage and stopping decisions.

\subsection{Requirement Answers after Full Disclosure}
\label{app:disclosed_understanding}
Requirement disclosure measures the extent to which the coding agent elicits complete information about initially hidden requirements through questions to the client; clarification score assesses the agent's understanding of task requirements through requirement questions. The former concerns acquiring the needed information, while the latter concerns understanding it accurately. We therefore examine whether agents correctly answer questions about requirements for which they have obtained complete information through clarification.

To distinguish information acquisition from requirement understanding, we pair disclosure annotations from normally completed clarification conversations with answer judgments from the same runs, without filtering outlying round counts. We select initially hidden RUs annotated as fully disclosed at the end of clarification. Table~\ref{tab:disclosed_understanding} reports Partial and Incorrect answer rates using the number of eligible RUs for each model as the denominator.

\begin{table}[htbp]
\centering\papertablestyle
\caption{Partial and Incorrect answers among fully disclosed hidden RUs. Counts give the eligible RUs per model; remaining answers are Correct.}
\label{tab:disclosed_understanding}
\begin{tabular}{@{}lrrr@{}}
\toprule
Model & Fully disclosed RUs & Partial (\%) & Incorrect (\%) \\
\midrule
GPT-6 Astra & 249 & 2.4 & 3.2 \\
Claude Fable 5.1 & 247 & 2.8 & 4.9 \\
Claude Opus 5 & 261 & 3.8 & 4.6 \\
GPT-5.6 Sol & 212 & 10.4 & 2.8 \\
Qwen-3.8 Max & 220 & 3.2 & 6.4 \\
GLM-5.3 & 202 & 4.0 & 6.4 \\
Kimi K3 & 198 & 8.1 & 5.1 \\
GPT-5.6 Luna & 201 & 14.9 & 3.0 \\
GLM-5.3 Flash & 172 & 4.7 & 2.9 \\
\bottomrule
\end{tabular}
\end{table}

Even after full disclosure, 2.8\%--6.4\% of answers are graded Incorrect and another 2.4\%--14.9\% are Partial across models. Successfully acquiring requirement information therefore does not guarantee accurate understanding, highlighting the value of independent requirement-answer assessment.

The two measures also differ in scope and scoring. Disclosure rate counts the proportion of initially hidden RUs for which the coding agent obtains complete information; clarification score covers all requirement questions and gives half credit to Partial answers. Agents can also reason from the supplied materials. The two percentages therefore should not be directly subtracted, and disclosure rate is not an upper bound on clarification score.

\subsection{Requirement Understanding and Implementation}
\label{app:joint_results}
\begin{figure}[!htb]
\centering
\includegraphics[width=0.90\linewidth]{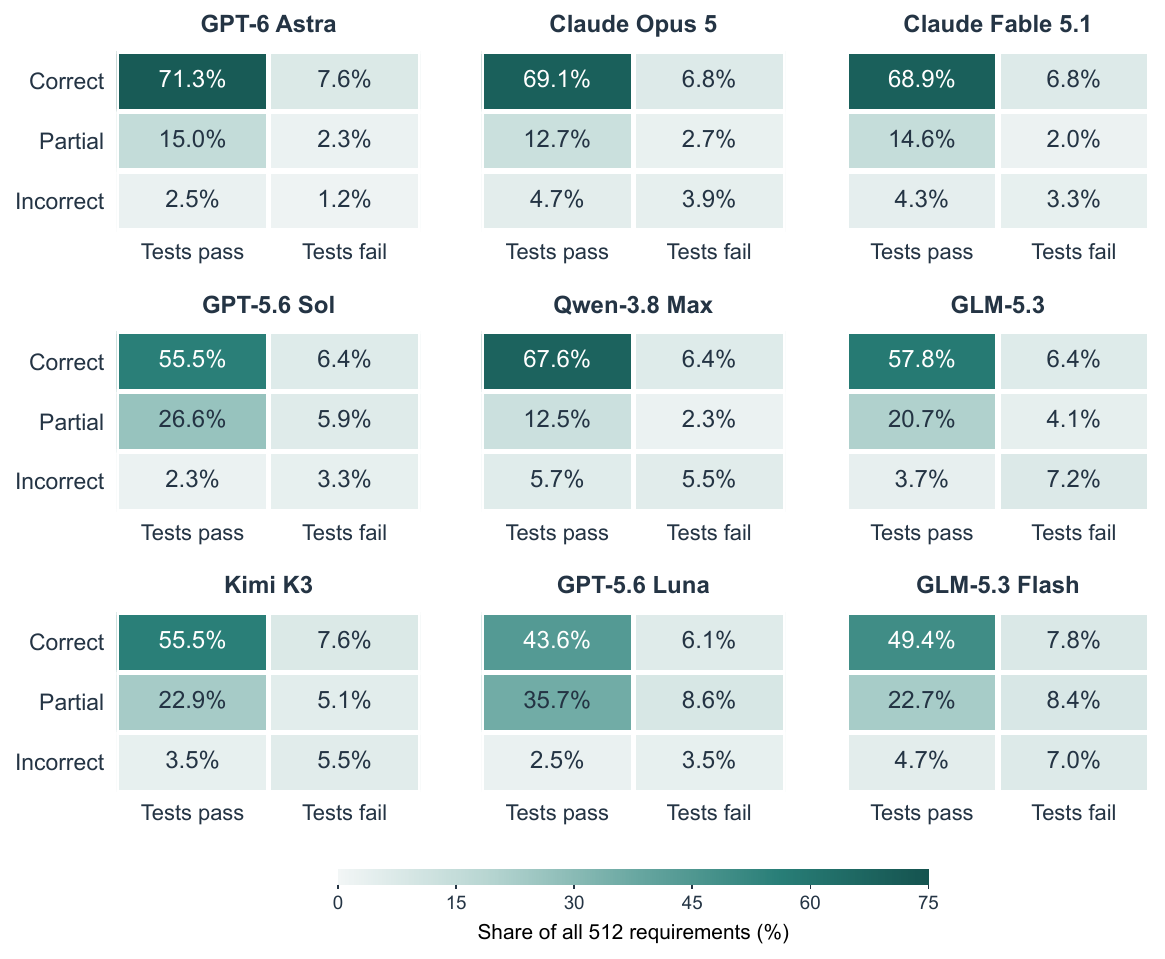}
\caption{Joint distribution of requirement answers and code-test outcomes. Cells show percentages of all 512 RUs per model; the six cells in each panel sum to 100\%, up to rounding. All panels share one color scale and follow the main-table model order. Tests pass and Tests fail refer to acceptance tests for the corresponding RU.}
\label{fig:joint_distribution}
\end{figure}

The strict answer-correctness rate gives credit only to fully correct answers and averages tasks equally. Answer-label counts and joint answer/code outcomes instead count individual RU observations. Table~\ref{tab:answer_breakdown} reports strict answer correctness and label counts, while Figure~\ref{fig:joint_distribution} reports the full distribution of answer and code outcomes.

\begin{table}[!htb]
\centering\papertablestyle
\caption{Strict answer correctness and answer-label counts over 100 tasks per model. Each model has 512 requirement answers.}
\label{tab:answer_breakdown}
\begin{tabular}{@{}lrrrr@{}}
\toprule
Model & Strict (\%) & Correct & Partial & Incorrect \\
\midrule
GPT-6 Astra & 79.0 & 404 & 89 & 19 \\
Claude Fable 5.1 & 76.1 & 388 & 85 & 39 \\
Claude Opus 5 & 76.3 & 389 & 79 & 44 \\
GPT-5.6 Sol & 62.2 & 317 & 166 & 29 \\
Qwen-3.8 Max & 74.0 & 379 & 76 & 57 \\
GLM-5.3 & 64.7 & 329 & 127 & 56 \\
Kimi K3 & 63.2 & 323 & 143 & 46 \\
GPT-5.6 Luna & 49.9 & 254 & 227 & 31 \\
GLM-5.3 Flash & 58.1 & 293 & 159 & 60 \\
\bottomrule
\end{tabular}
\end{table}

Partial answers outnumber Incorrect answers for every model, indicating that incomplete answers are more common than answers graded incorrect. The three-level rubric distinguishes partial from complete knowledge, while strict correctness reveals whether agents can fully understand and articulate requirements.

Figure~\ref{fig:joint_distribution} complements Section~\ref{sec:rq3} with the full joint distribution of three answer labels and two code-test outcomes. Whereas Figure~\ref{fig:agreement} compares test-pass rates within answer categories, each cell here uses all 512 RUs for the model as its denominator, showing the prevalence of each outcome in the evaluation.

\paragraph{Successful software delivery requires both clarification and implementation capabilities.} Requirement clarification and code implementation are related. As shown in Section~\ref{sec:rq3}, requirements answered correctly have higher implementation success rates than those answered partially or incorrectly across all nine models. The full joint distribution in Figure~\ref{fig:joint_distribution} further shows that the two capabilities do not always align: agents may correctly understand and articulate requirements yet fail to implement them, or pass the corresponding tests without fully demonstrating their understanding in the answers. Understanding established during clarification thus provides an important foundation for implementation, while delivery also depends on the agent's ability to translate requirements into code.

\paragraph{Agent development should improve both clarification quality and the translation of requirements into code.} One possible explanation for incomplete answers accompanying successful implementations is that agents further infer the required behavior during coding by combining available materials, implementation details, and their own checks. Conversely, correct answers accompanying failed implementations suggest that accurate requirement understanding may be lost or insufficiently translated into code. These observations motivate carrying clarified requirements through implementation planning and verification, and studying how agents identify and resolve gaps in understanding during coding. Beyond clarifying what is required, agents should consistently translate that understanding into verifiable software behavior.

\subsubsection{Answer--Implementation Cases}
\label{app:cases}
Task 015 provides a simple example: when a filter matches no items, the application should return a valid empty result rather than an error. GPT-5.6 Luna correctly describes the filtering rules in its answer to RU-003 but omits the behavior when no items match, receiving a Partial label. Its implementation from the same run passes the corresponding acceptance test, which includes inputs with no matches and requires HTTP 200 with an empty result list.

Task 053 illustrates the reverse. In its Correct-rated answer to RU-001, GPT-5.6 Luna states that malformed historical data should raise \texttt{ValueError}, whereas a valid empty history should return empty results. The implementation, however, first converts the input into a list, turning an invalid empty dictionary into a valid empty list and returning empty results without an error. The acceptance log from the same run records failure on this input, and inspection of the implementation confirms the conversion logic.

Together, the cases show why requirement answers and implementation outcomes are not interchangeable: task 015 implements a behavior omitted from the answer, while task 053 fails to enforce an input rule correctly stated in the answer. They motivate translating clarified requirements into concrete checks, explicitly distinguishing boundary cases such as a valid empty result from an invalid input, and checking code behavior against those requirements throughout implementation.

\FloatBarrier
\section{Clarification Timing}
\label{app:timing}
Should clarification be completed before coding or continue during development? We compare two workflows using GPT-5.6 Luna and GLM-5.3 Flash on the same 100 tasks per model, with the same scoring rubric as the main comparison.

\textbf{Clarification before coding} follows the standard protocol: the coding agent inspects materials and asks the client questions, autonomously ends clarification, and then implements from the frozen clarification context without further client interaction. \textbf{Continued clarification} allows the agent to alternate between questions and coding, retaining existing code while obtaining further requirements during development. The distinction is whether client interaction ends before implementation or can be interleaved with it.

\begin{table}[htbp]
\centering\papertablestyle
\caption{Implementation and clarification scores (\%) for two workflows on 100 tasks per model. Clarification scores measure requirement-answer performance.}
\label{tab:clarification_timing}
\begin{tabular}{@{}lrrrr@{}}
\toprule
Model & \multicolumn{2}{c}{Clarification before coding} & \multicolumn{2}{c}{Continued clarification} \\
\cmidrule(lr){2-3}\cmidrule(lr){4-5}
 & \shortstack{Implementation\\score} & \shortstack{Clarification\\score} & \shortstack{Implementation\\score} & \shortstack{Clarification\\score} \\
\midrule
GPT-5.6 Luna & 49.0 & 71.8 & 26.0 & 59.0 \\
GLM-5.3 Flash & 39.0 & 73.2 & 13.0 & 50.9 \\
\bottomrule
\end{tabular}
\end{table}

\textbf{Clarification before coding achieves higher implementation and clarification scores for both models.} Relative to continued clarification, implementation scores are higher by 23 and 26 percentage points for Luna and Flash, respectively, and clarification scores are higher by 12.8 and 22.3 points (Table~\ref{tab:clarification_timing}). Paired outcomes show the same pattern: 30 and 29 tasks pass only with clarification before coding, compared with seven and three only with continued clarification. In these workflow comparisons, allowing further questions during coding does not yield more accurate requirement answers or higher delivery success.

One possible explanation is that newly acquired information during continued clarification may repeatedly change the agent's earlier understanding and decisions. When client replies, prior assumptions, and code revisions are interleaved during development, the agent may confuse superseded assumptions with current requirements and struggle to maintain a coherent understanding. The lower final clarification scores for both models provide indirect support for this hypothesis: additional opportunities for interaction do not automatically yield more accurate requirement understanding; effectively integrating evolving information also matters.

These results motivate improving how interleaved agents integrate new information into both their requirement understanding and existing code. Agents could check that critical requirements are clear before coding and explicitly update requirement records, implementation plans, and relevant checks when later clarification changes an assumption. Such coordination can help turn continued interaction into consistent software behavior.

\FloatBarrier
\section{Effect of Supervisor Review}
\label{sec:validity}
The Supervisor reviews client replies and provides reasons for requesting revisions, which the Client uses to revise its responses. We compare the first draft with the final approved reply to examine how review changes reply quality. Saved conversations contain 4,418 reviewed questions, of which 561 receive at least one revision request. We assess the correctness of the first draft and final reply against the task references, with manual verification where needed.

Among the 561 revised questions, 46.0\% of replies change from incorrect to correct, while 2.9\% change from correct to incorrect (Table~\ref{tab:supervisor_effect}). The former represents successful correction; the latter represents harmful revision that introduces an error. The other two outcomes indicate preserved correctness and failure to correct an existing error, respectively.

\begin{table}[htbp]
\centering\papertablestyle
\caption{First-draft to final-reply outcomes among the 561 revised questions. Values are percentages and sum to 100\%, up to rounding.}
\label{tab:supervisor_effect}
\begin{tabular}{@{}lr@{}}
\toprule
Transition & Percentage (\%) \\
\midrule
Incorrect $\rightarrow$ correct & 46.0 \\
Correct $\rightarrow$ incorrect & 2.9 \\
Correct $\rightarrow$ correct & 43.7 \\
Incorrect $\rightarrow$ incorrect & 7.5 \\
\bottomrule
\end{tabular}
\end{table}

Successful corrections substantially outnumber harmful revisions, indicating a net accuracy benefit among replies selected for revision. This supports reviewing client replies before delivery to the coding agent to reduce erroneous information entering the subsequent clarification context.

\section{System Prompts}
\label{app:prompts}
The following boxes reproduce the system prompts used for clarification, client simulation, reply review, requirement answering, implementation, and grading. Line wrapping is adjusted for readability. Tool definitions and task-specific context accompany these prompts at runtime. We also include the instruction appended to the coding prompt.

\subsection{Clarification Agent}
The clarification agent inspects available materials, asks focused questions, and decides when to stop, without implementing during this phase.
\systemprompt{P1: Clarification agent}{clarification_system}

\subsection{Client}
The client answers requested business facts and delivery agreements without automatically disclosing the full specification.
\systemprompt{P2: Client}{client_system}

\subsection{Supervisor}
The supervisor reviews factuality and scope before the candidate sees a reply. Internal feedback is not passed to the candidate, and disclosure annotations are not a prerequisite for approval.
\systemprompt{P3: Supervisor}{supervisor_system}

\subsection{Requirement-Answer Branch}
The answer branch receives the frozen context and responds to evaluation questions. Its answers do not enter the coding branch.
\systemprompt{P4: Requirement-answer branch}{audit_system}

\subsection{Coding Branch}
The coding branch implements the deliverable from its independent context, without receiving the other branch's answers or scores.
\systemprompt{P5: Coding branch base prompt}{coding_system}

\subsection{Coding Runtime Instruction}
The following instruction is appended to the coding system message and specifies the workspace and native tools; it is not a client requirement.
\systemprompt{P5a: Coding runtime addition}{coding_runtime_addition}

\subsection{Judge Agent}
The main comparison, information-condition controls, and clarification-timing comparison use the following prompt to grade requirement answers as Correct, Partial, or Incorrect.
\systemprompt{P6: Judge agent}{tiered_audit_judge_system}